\documentclass[twocolumn]{aastex701}

\usepackage{amsmath}

\graphicspath{{Figures/}{./}}

\begin{document}

\title{TeV Gamma Rays and Neutrinos from Winds Driven by Radiatively Inefficient Accretion Flows of Low-Luminosity Active Galactic Nuclei}

\author[orcid=0009-0004-5978-1785,gname='Nobuyuki',sname='Sakai']{Nobuyuki Sakai}
\affiliation{Department of Earth and Space Science, Graduate School of Science, The University of Osaka, Toyonaka, Osaka 560-0043, Japan}
\affiliation{Center for Cosmology and AstroParticle Physics (CCAPP), The Ohio State University, Columbus, OH 43210, USA}
\email[show]{u938638f@ecs.osaka-u.ac.jp}  

\author[orcid=0000-0002-0005-2631,gname='John',sname='Beacom']{John F. Beacom} 
\affiliation{Center for Cosmology and AstroParticle Physics (CCAPP), The Ohio State University, Columbus, OH 43210, USA}
\affiliation{Department of Physics, The Ohio State University, Columbus, OH 43210, USA}
\affiliation{Department of Astronomy, The Ohio State University, Columbus, OH 43210, USA}
\email{beacom.7@osu.edu}

\author[orcid=0000-0002-5358-5642,gname='Kohta',sname='Murase']{Kohta Murase} 
\affiliation{Department of Physics, The Pennsylvania State University, University Park, PA 16802, USA}
\affiliation{Institute for Gravitation and the Cosmos, The Pennsylvania State University, University Park, PA 16802, USA}
\affiliation{Department of Astronomy \& Astrophysics, The Pennsylvania State University, University Park, PA 16802, USA}
\affiliation{Center for Gravitational Physics and Quantum Information, Yukawa Institute for Theoretical Physics, Kyoto University, Kyoto 606-8502, Japan}
\email{murase@psu.edu}


\begin{abstract}

Recently, the Large High Altitude Air Shower Observatory (LHAASO) reported the first TeV gamma-ray detection from a low-luminosity active galactic nucleus (LLAGN) lacking a strong jet. 
This source, NGC 4278, is one of the nearest LLAGNs that can be studied in detail as a template for more distant objects.
Most previous studies have focused on the month-scale-variable gamma-ray activities, seeking to explain them with weak jets.
We focus instead on NGC~4278's quiet-period emission, proposing a different mechanism based on a wind launched from a radiatively inefficient accretion flow (RIAF).
In particular, we model particle acceleration at a wind-driven shock and calculate the resulting multimessenger emission. 
To match the TeV gamma-ray fluxes reported by LHAASO in the quiet period, our model requires a wind power of $\lesssim2\times10^{43}~\mathrm{erg~s^{-1}}$ with a spectral index of injected cosmic rays of 2.0.  
We show that this is achievable through Bondi accretion of the ionized gas observed on the 100-pc scale, for a shallow accretion profile. 
We discuss the testability of our wind-driven shock scenario with multimessenger observations.
As another important application, we extend our model to Sgr~A*, the nearest LLAGN, and show that its TeV gamma-ray emission is reproduced with reasonable parameters. 
Our results suggest that RIAF-driven winds may be a generic emission mechanism in LLAGNs, making this class a potentially important population of both gamma-ray emitters and cosmic-ray accelerators.

\end{abstract}



\section{Introduction}
\label{sec:introduction} 

Though bright diffuse extragalactic backgrounds of high-energy particles --- cosmic rays (CRs), gamma rays, and neutrinos --- have long been observed, their sources remain uncertain.
Gamma rays and neutrinos have been observed to over $10^{15}$~eV \citep[e.g.,][respectively]{2023PhRvL.131o1001C, 2026PhRvL.136l1002A} and CRs have been observed to over $10^{20}$~eV \citep[e.g.,][]{2008PhRvL.101f1101A, 2024PhRvL.133d1001A}. 
Which source classes dominate the contributions of these particles to their diffuse backgrounds?  
Progress requires confronting theoretical models with multiwavelength and multimessenger observations of individual sources.

Active galactic nuclei (AGNs) are an important type of CR accelerator, though it is unknown if they primarily accelerate protons or electrons.
AGNs are bright regions in the central regions of galaxies, caused by gas accretion onto supermassive black holes \citep[SMBHs, see e.g.,][for reviews]{1984ARA&A..22..471R, 2008bhad.book.....K}. 
As a source class, AGNs are important to understand because of their commonality and diversity. 
Bolometric luminosities of nearby AGNs span $\sim10^{38}$--$10^{44}~\mathrm{erg~s^{-1}}$ \citep{2008ARA&A..46..475H}.
The advent of sensitive new detectors has opened a window into high-energy processes in AGNs. 
In the GeV gamma-ray sky, the Large Area Telescope on board the \textit{Fermi} space gamma-ray telescope (\textit{Fermi}-LAT) has detected more than 3000 AGNs \citep[4FGL-DR4 catalog,][]{2023arXiv230712546B}.
The IceCube Neutrino Observatory \citep{2017JInst..12P3012A} has reported possible signals of neutrino emission from some AGNs \citep[e.g.,][for NGC~1068 and NGC~4151, respectively]{2022Sci...378..538I, 2025ApJ...988..141A}.

By numbers, low-luminosity AGNs (LLAGNs) dominate the AGN population. LLAGNs typically have luminosities below $10^{41}~\mathrm{erg~s^{-1}}$. LLAGNs are abundant, occurring in roughly 30\% of nearby galaxies \citep[e.g.,][]{2008ARA&A..46..475H}. 
Interestingly, LLAGNs are not just faint but have different physics from bright AGNs like Seyferts. 
LLAGNs are characterized by radiatively inefficient accretion flows (RIAFs), truncated geometrically thin disks, and jets/winds from RIAFs \citep[e.g.,][]{1977ApJ...214..840I, 1994ApJ...428L..13N,1999MNRAS.303L...1B,2008ARA&A..46..475H,2014ARA&A..52..529Y}.

The first source catalog of the Large High Altitude Air Shower Observatory \citep[LHAASO,][]{2016NPPP..279..166D, 2019arXiv190502773C} reported the detection of a very high-energy gamma-ray source, 1LHAASO~J1219+2915 \citep{2024ApJS..271...25C}.
\citet{2024ApJ...971L..45C} further investigated this point-like source in detail and reported the first detection of an LLAGN without a strong jet. 
The source is NGC~4278, a low-ionization nuclear emission-line region (LINER) type LLAGN located at a distance of $D\approx16~\mathrm{Mpc}$ \citep{2001ApJ...546..681T}.
Their variability analysis divided the TeV emission into an active period and a quiet period, and found that the flux increased by a factor of $\approx 7$ on timescales of a month or longer (see Section~\ref{sec:NGC4278 gamma} for details).
Importantly, the source is not spatially resolved in gamma rays, preventing constraints on the size of the emission region from gamma-ray observations alone.

If the TeV emission is associated with NGC 4278, it indicates efficient CR acceleration in or around this LLAGN.
This motivates efforts to understand the origin of TeV photons and, by extension, CR acceleration in the LLAGN class of sources.
Hereafter, we consider that the TeV emission originates from NGC~4278.
The physical origin of the gamma-ray emission in NGC~4278 remains under debate.
Weak, compact jets \citep[][all focus on the active-period emission and the last one
also focuses on the quiet-period emission]{2024ApJ...971L..45C, 2024ApJ...974...56D,
2024ApJ...974..134L, 2024ApJ...977L..16B, 2026ApJ..1007...10C} and an accretion flow
\citep[][focusing on the quiet-period emission]{2025PASJ..tmp..127S} have been proposed
as candidate mechanisms.
In addition, \citet{2026ApJ..1003...71Y} considered both a jet and a wind, modeling the active and
quiet periods.
\citet{Das_2026} show that gamma rays cannot escape from an RIAF due to internal absorption through the two-photon annihilation. 
Thus, the active-period emission is more likely to come from outflows, and the observed TeV luminosity is consistent with the jet power \citep{2026ApJ..1007...10C}. 
On the other hand, the quiet-period TeV emission is less demanding for energetics requirements. 
For example, \citet{2025PASJ..tmp..127S} show that only if the AGN activity had been at least ten times higher in the past than at present, CR protons escaping from an RIAF could account for the TeV emission. 

In this paper, we focus on the quiet-period emission and explore a scenario where the gamma-ray emission originates from an RIAF-driven wind. 
Generally, LLAGNs are thought to host RIAFs, and they naturally produce powerful wide-angle winds \citep[e.g.,][]{1999MNRAS.303L...1B, 2008ARA&A..46..475H, 2014ARA&A..52..529Y} in addition to collimated jets.
RIAF winds can be possible CR accelerators as an analogy of disk-driven winds in more luminous Seyfert galaxies \citep[e.g.,][]{Tamborra2014,2016A&A...596A..68L,Liu2018,2022arXiv220702097I, 2023MNRAS.526..181P, 2025ApJ...980..131S, 2025JCAP...07..013P, 2026arXiv260617490S}.
Although \citet{2026ApJ..1003...71Y} suggest that the TeV flux can be explained by leptohadronic emission from winds, the wind energetics are not explicitly presented, and thus, it remains uncertain whether winds in the LLAGN can account for the TeV emission during the quiet period. 
Considering gas accretion as the energy budget of the wind, we also hypothesize that the TeV emission originates from CRs accelerated at a shock generated by an RIAF-driven wind. 
Compared to \citet{2026ApJ..1003...71Y}, we consider a larger-scale shock suggested by the optical observation \citep{2024A&A...683A..43H} and discuss observed multiphase gas as the wind-energy budget in more detail.
Also, our work qualitatively differs from both \citet{Das_2026} that focus on RIAF X-ray emission without explaining TeV gamma-ray data and \citet{2026ApJ..1007...10C} that investigate TeV emission from jets using RIAF photons as a target radiation field. 
To generalize the importance of LLAGNs in gamma-ray production and CR acceleration, we extend our model to Sagittarius~A* (Sgr~A*), another LLAGN at the Galactic center, from which TeV gamma rays have been detected \citep[e.g.,][]{2016Natur.531..476H}.
Our results have implications for the potential contribution of the LLAGN source class to diffuse high-energy backgrounds.

This paper is organized as follows.
First, we summarize the observed properties of NGC~4278 in Section~\ref{sec:NGC4278}. 
In Section~\ref{sec:method}, we construct an emission model based on an RIAF-driven wind. 
Section~\ref{sec:results} presents spectral energy distributions (SEDs), consistent with the TeV observations.  
In Section~\ref{sec:discussion}, we explain whether the required physical parameters are plausible in realistic LLAGN environments and how to constrain our model and other models observationally.
We apply our model to Sgr~A* in Section~\ref{sec:Sgr A*}.
Finally, we conclude in Section~\ref{sec:conclusion}.


\section{Review of NGC 4278 Properties}
\label{sec:NGC4278}

In this section, we summarize the observational properties of NGC~4278.

\subsection{Multiwavelength Properties}
\label{sec:NGC4278 MWL}

NGC~4278 is an early-type galaxy classified as a LINER \citep[][]{1980A&A....87..152H} and is known to host an LLAGN \citep[e.g.,][]{1997ApJS..112..391H, 2010A&A...517A..33Y}.
The SMBH mass estimated by the $M_\mathrm{BH}$-$\sigma$ relation \citep{2002ApJ...574..740T} is $M_\mathrm{BH}=(3.90\pm0.54)\times10^8~M_\odot$ \citep{2003MNRAS.340..793W, 2005ApJ...625..716C, 2010A&A...517A..33Y} and the distance is $D\approx16~\mathrm{Mpc}$ \citep{2001ApJ...546..681T}. 
The bolometric luminosity estimated by integrating the SED from radio to X-ray is $\sim2$--$3\times10^{41}~\mathrm{erg~s^{-1}}$ \citep[][]{2010A&A...517A..33Y}, corresponding to $\sim10^{-5}L_\mathrm{Edd}$, where $L_\mathrm{Edd}=4\pi GM_\mathrm{BH}m_pc/\sigma_\mathrm{T}$ is the Eddington luminosity with the gravitational constant $G$, the proton mass $m_p$, the speed of light $c$, and the Thomson scattering cross section $\sigma_\mathrm{T}$.

Radio observations have revealed jet activities in this AGN.
Using the Very Long Baseline Array (VLBA) and a single antenna of the Very Large Array (VLA), \citet{2005ApJ...622..178G} revealed a two-sided structure with symmetric S-shaped jets emerging from a flat-spectrum core.
Its apparent velocity is $\lesssim0.2c$ \citep{2005ApJ...622..178G} and the estimated power of the jet is $\sim10^{42}~\mathrm{erg~s^{-1}}$ \citep{2012ApJ...758...65B}.

Multiwavelength observations have revealed three phases of gas in NGC~4278: warm ionized gas, hot ionized gas, and cold molecular gas.
These gas phases can be accreted by the SMBH and can supply the energy budget of the kinetic powers of RIAF winds.
In Section~\ref{sec:gas accretion}, we estimate kinetic powers of RIAF winds based on these three gas phases and discuss the possibilities of their being connected to the TeV gamma rays. 
In short, we will find that warm ionized gas is favored for this, but here we summarize observations of all the gas phases.

First, NGC~4278 hosts an outflow of warm ionized gas.
The observation by the Multi-Espectrógrafo en GTC de Alta Resolución para Astronomía \citep[MEGARA,][the field of view~$\sim200$~pc]{2024A&A...683A..43H}
reported signatures of an outflow with a shell-like morphology, a velocity of $\approx241~\mathrm{km~s^{-1}}$, and a spatial extent of $\approx122$~pc, based on their analysis of H$\alpha$ and [O~III] emission lines. 
They placed a lower limit on the kinetic power of $10^{38}~\mathrm{erg~s^{-1}}$ and estimated an electron density of $\sim300~\mathrm{cm^{-3}}$ assuming a temperature of $10^{4}$~K. 

Second, X-ray observations have revealed hot gas in this galaxy.
The temperature of the hot plasma ranges from $\approx0.2$ to $0.8$~keV \citep{2009A&A...506.1107G, 2010A&A...517A..33Y, 2012ApJ...758...65B, 2012ApJ...758...94P}. 
\citet{2012ApJ...758...65B} detected hot X-ray gas with a temperature of 0.35~keV and a bipolar morphology, indicating that it may be outflowing from the galaxy. 
The typical densities of this X-ray emitting gas are $\sim0.01$--$0.1~\mathrm{cm^{-3}}$ \citep{2012ApJ...758...65B, 2012ApJ...758...94P}.

Finally, this galaxy contains cold molecular gas.
Atacama Large Millimeter/submillimeter Array (ALMA) observations \citep{2025PASJ..tmp..127S} showed that NGC~4278 hosts a massive molecular cloud surrounding the nucleus, with a mass of $\sim10^7~M_\odot$ and a size of $\sim100$~pc, corresponding to a number density of $\sim10^2~\mathrm{cm^{-3}}$.
\citet{2011MNRAS.414.1827T} also revealed H$_2$ gas with temperatures of $\sim100$--1000~K.


\subsection{Gamma-ray Wavelengths}\label{sec:NGC4278 gamma}
 
\citet{2024ApJ...971L..45C} divided the LHAASO observations of TeV emission into active and quiet periods.
During the active period, an observed flux of $7.0 \times 10^{-13}~\mathrm{photons~cm^{-2}~s^{-1}}$ was reported in the 1--10~TeV band, with a spectral index of 2.56.
The 1--10~TeV flux during the quiet period is expected to be $\sim 10^{-13}~\mathrm{photons~cm^{-2}~s^{-1}}$, although the quiet-period flux is not explicitly presented in \citet{2024ApJ...971L..45C}.
They also reported the gamma-ray luminosity of $L_\mathrm{0.1-10~TeV}\approx3.0\times10^{41}~\mathrm{erg~s^{-1}}$.

Other TeV observations of NGC~4278 have reported no significant emission.
VERITAS observations of NGC~4278 \citep{VERITAS_2026} reported a nondetection.
A 95\% confidence-level differential upper limit of $8.3\times10^{-9}~\mathrm{TeV^{-1}~m^{-2}~s^{-1}}$ at $680$~GeV, corresponding to an energy flux of $\approx6\times10^{-13}~\mathrm{erg~cm^{-2}~s^{-1}}$, was obtained for the quiet state.
Combining this limit with \textit{Fermi}-LAT and LHAASO data, they inferred that the peak of the high-energy SED component lies between $\sim100$~GeV and $\sim2$~TeV.
The MAGIC Collaboration \citep{MAGIC_2026} reanalyzed archival observations, none of which overlap the LHAASO active period, and likewise reported a nondetection.

The GeV gamma-ray observations are not fully consistent among analyses. 
First, the \textit{Fermi}-LAT 4FGL-DR4 catalog \citep{2023arXiv230712546B} does not note NGC~4278.
\citet{2024ApJS..271...10W,2024ApJ...974..134L} analyzed \textit{Fermi}-LAT data in time periods that include the LHAASO-reported active period, and they found no significant GeV emission. 
In contrast, \citet{2024ApJ...977L..16B} reported a GeV gamma-ray detection from NGC~4278 with a significance of $\sim4.3\sigma$ using \textit{Fermi}-LAT data restricted to the time interval of the first LHAASO campaign.
More recently, \citet{VERITAS_2026,2026ApJ..1007...10C} analyzed the active and quiet periods separately, and found no significant emission in either.
The quiet-period upper limits of \citet{VERITAS_2026} reach an energy flux of $\approx2\times10^{-13}~\mathrm{erg~cm^{-2}~s^{-1}}$ at a few GeV.


\section{Wind-Emission Model}
\label{sec:method}

In this section, we build a model that describes the dynamics of an RIAF-driven wind and wind-driven shock (Section~\ref{sec:model dynamics}) and gamma-ray and neutrino emission from CR electrons and protons accelerated at the shock (Section~\ref{sec:model emission}). 

\subsection{RIAF Wind and Shock Dynamics}\label{sec:model dynamics}

We consider a wind from an RIAF with a kinetic power of $\dot{K}_\mathrm{w}$.
An RIAF wind is thought to be magnetically and/or thermally driven
and its energy budget is likely due to Bondi accretion \citep[e.g.,][]{1999MNRAS.303L...1B,2003ApJ...598..301Y,2014ARA&A..52..529Y}.
One necessary condition is that the wind kinetic power, which is the energy budget of CR acceleration and gamma-ray emission, must significantly exceed the observed gamma-ray luminosity. 

The interaction of the RIAF wind with the interstellar medium (ISM) produces strong shock waves, which can accelerate charged particles.
We parameterize the ISM density as $n_0$ with a uniform, spherically symmetric distribution motivated by the ALMA observation of NGC~4278 \citep{2025PASJ..tmp..127S}. 
The forward-shock (FS) radius follows a self-similar solution \citep[$R_\mathrm{FS}\propto t_\mathrm{w}^{3/5}$, where $t_\mathrm{w}$ is the wind age, e.g.,][]{2012MNRAS.425..605F} and subsequently, the forward-shock velocity can be related to $R_\mathrm{FS}$ as
\citep{2012MNRAS.425..605F, 2024ApJ...968..116Y}, 
\begin{align}\label{eq:shock velocity}
    V_\mathrm{FS}=\frac{1}{2}\left(\frac{18\dot{K}_\mathrm{w}}{5\pi n_0m_p}\right)^{1/3}R_\mathrm{FS}^{-2/3}.
\end{align}
Then, the kinetic power of the forward shock is given by
\begin{align}\label{eq:shock power}
    \dot{K}_\mathrm{FS}=2\pi R_\mathrm{FS}^2n_0m_pV_\mathrm{FS}^3=\frac{9}{10}\dot{K}_\mathrm{w}.
\end{align}
This quantity is used as an energy budget of CR injection power, motivated by particle-in-cell simulations of non-relativistic collisionless shocks \citep[e.g.,][]{2014ApJ...783...91C, 2015PhRvL.114h5003P}.

\subsection{CR Acceleration and Multimessenger Emission}\label{sec:model emission}

Wind-driven shocks accelerate charged particles to relativistic energies through diffusive shock acceleration \citep[e.g.,][]{1978MNRAS.182..147B,1978MNRAS.182..443B,1978ApJ...221L..29B,1983RPPh...46..973D}. 
The DSA rate at a parallel shock for a CR particle with an energy $E_\mathrm{CR}$ is approximated as \citep{1983RPPh...46..973D}
\begin{align}\label{eq:DSA timescale}
    \tau_\mathrm{DSA}^{-1}(E_\mathrm{CR})=\frac{3eB_\mathrm{ISM}V_\mathrm{FS}^2}{20\xi_Bc}E_\mathrm{CR}^{-1},
\end{align}
where we consider Bohm-type diffusion, $e$ is the elementary charge, $B_\mathrm{ISM}$ is the magnetic field in the upstream ISM, and $\xi_B\equiv {(B/\delta B)}^2$ is the Bohm factor. 
In this work, we fix $\xi_B=1$ as a benchmark for simplicity (but see Section~\ref{sec:test} for a more general case).
We use the same field for the synchrotron emission from the shocked region, so $B_\mathrm{ISM}$ should be understood as an effective field that characterizes both the acceleration site and the radiating region.
Although shock compression alone would make the downstream field larger by a factor of $\approx4$, CR-driven instabilities amplify the field in the upstream precursor by orders of magnitude \citep[e.g.,][]{2004MNRAS.353..550B}, which reduces the contrast between the two regions.
We therefore set the shocked magnetic field to $B_\mathrm{ISM}$ for simplicity, implicitly including the amplification in the quoted values.

For the energy distribution of the CR injection rate, we assume a power-law spectrum with index $q_\mathrm{CR}$ and an exponential cutoff.
The normalization of the injection spectrum of CRs is determined by the fact that an energy fraction $\epsilon_{p/e}$ of the shock power given by Equation~(\ref{eq:shock power}) is converted to the total CR proton/electron power.
Unless otherwise noted, we adopt $\epsilon_{p}=0.1$ inferred from observations of supernova remnants \citep[][]{2013Sci...339..807A} and particle-in-cell simulations \citep[e.g.,][]{2014ApJ...783...91C,2026ApJ..1004...27L}.
The electron-to-proton ratio $\epsilon_e/\epsilon_p$ is less certain, and both particle-in-cell simulations and supernova-remnant observations indicate that it depends on the shock velocity and infer a wide range of $\sim10^{-4}$--$10^{-1}$ \citep[e.g.,][]{2012A&A...538A..81M,2015PhRvL.114h5003P,2017MNRAS.464.2326S,2021MNRAS.508.6142M,2026ApJ..1004...27L}.
We therefore treat $\epsilon_e/\epsilon_p$ as a model parameter.

We determine the maximum proton/electron energy $E_{p/e,\mathrm{cut}}$ by the balance between the acceleration rate given by Equation~(\ref{eq:DSA timescale}) and the larger rate among the advection rate 
\begin{align}\label{eq:advection timescale}
    \tau_\mathrm{adv}^{-1}=V_\mathrm{FS}/R_\mathrm{FS}    
\end{align}
and the energy-loss rate of CRs.
We consider the energy loss via synchrotron and inverse-Compton (IC) scattering for electrons \citep[see Equations~15 and 16 in][for these loss rates]{2023MNRAS.524...76Z}.
The spectrum of seed photons for IC scattering depends on the sources.
CR protons lose their energy through inelastic \textit{pp} interactions, with the energy-loss rate 
\begin{align}
    \tau_{pp}^{-1}\sim\kappa_{pp}n_0\sigma_{pp}c,
\end{align}
where $\kappa_{pp}\approx0.5$ is the inelasticity.
The \textit{pp} cross section $\sigma_{pp}$ is adopted from \citet[][see also \citealt{2025PhRvD.112d3009G}]{2014PhRvD..90l3014K}.

An injected CR distribution approaches a steady state by escaping from the wind system or losing energy. 
To solve the series of coupled transport equations in steady-state for various particles, including photons, electrons, neutrinos, neutrons, and protons, we use the Astrophysical Multimessenger Emission Simulator \citep[\texttt{AMES}, see][for details]{2023MNRAS.524...76Z,Murase2024}.
We account for advection escape and multiple energy-loss processes, including synchrotron and IC scattering for electrons and \textit{pp} interactions, photomeson production, and the Bethe-Heitler process.
Since charged particles escape by advection in our model, the fraction of CR proton energy converted into pions in \textit{pp} interactions is $f_{pp}=\min[1,\tau_{pp}^{-1}/\tau_\mathrm{adv}^{-1}]$.

CR particles in a steady state produce multimessengers, including photons and neutrinos.
We consider \textit{pp} interactions, synchrotron and IC scattering from primary accelerated electrons and secondary co-produced electrons by \textit{pp} interactions as emission processes using \texttt{AMES}.
We set the density of cold target protons for \textit{pp} interactions to $n_0$ rather than to the compressed downstream value.
This is appropriate for a one-zone treatment: shock compression raises the downstream density to $\approx4n_0$ but reduces the width of the shocked shell by the same factor through mass conservation, so that the target column traversed by a CR proton before it is advected out of the system is essentially unchanged.

Finally, gamma rays are attenuated by two-photon annihilation with the extragalactic background light (EBL) and the cosmic microwave background during propagation to Earth.
We adopt the EBL model of \citet{2012MNRAS.422.3189G}, following \texttt{AMES}, which gives $\tau_{\gamma\gamma}=1$ at $\approx36$~TeV for the distance of NGC~4278.
The attenuation is at most $\sim10\%$ across the LHAASO band, but it suppresses the flux by more than two orders of magnitude at $\gtrsim100$~TeV.
We include the factor $e^{-\tau_{\gamma\gamma}}$ in all gamma-ray spectra presented below.

\subsection{Strategies for Reproducing the TeV Emission in NGC~4278}\label{sec:stratege}

We aim to reproduce the TeV gamma-ray flux from NGC~4278 obtained by LHAASO with our wind model, focusing on the quiet period. 
The LHAASO-confirmed active and quiet periods may originate from different physical regions: a compact component may dominate during the active state, while a large-scale wind shock may provide the persistent baseline TeV emission.
In this work, we focus on the quiet period because an RIAF wind with a $\sim100$~pc-scale inferred by \citet{2024A&A...683A..43H} cannot explain the observed $\sim100$~day-scale variability during the active period \citep[$\ll 100~\mathrm{pc}/c$,][]{2024ApJ...971L..45C}. 
In addition, emission during the quiet period may be more important for diffuse fluxes if active phases are rare, as in NGC~4278.

We consider a leptohadronic scenario and a hadronic scenario, in which the sum of IC scattering and \textit{pp} interactions, or only \textit{pp} interactions, contribute the TeV band, respectively.
Here, we treat $\dot{K}_\mathrm{w}$, $R_\mathrm{FS}$, $B_\mathrm{ISM}$, $q_\mathrm{CR}$,
$n_0$, and $\epsilon_e/\epsilon_p$ as free parameters to reproduce the TeV gamma-ray flux
in the quiet period without significantly overshooting observed radio fluxes.
The quiet-period \textit{Fermi}-LAT upper limits of \citet{VERITAS_2026} were not used in
fixing the benchmarks, and we compare our models with them in
Sections~\ref{sec:leptohadronic} and \ref{sec:test}.
We fix the other model parameters to the fiducial values; $\epsilon_p=0.1$ and $\xi_B=1$.
We treat all multiwavelength fluxes collected from \citet{https://doi.org/10.26132/ned1} as upper limits.

\begin{figure*}
\centering
    \begin{minipage}{0.48\linewidth}
        \centering
        \includegraphics[width=\linewidth]{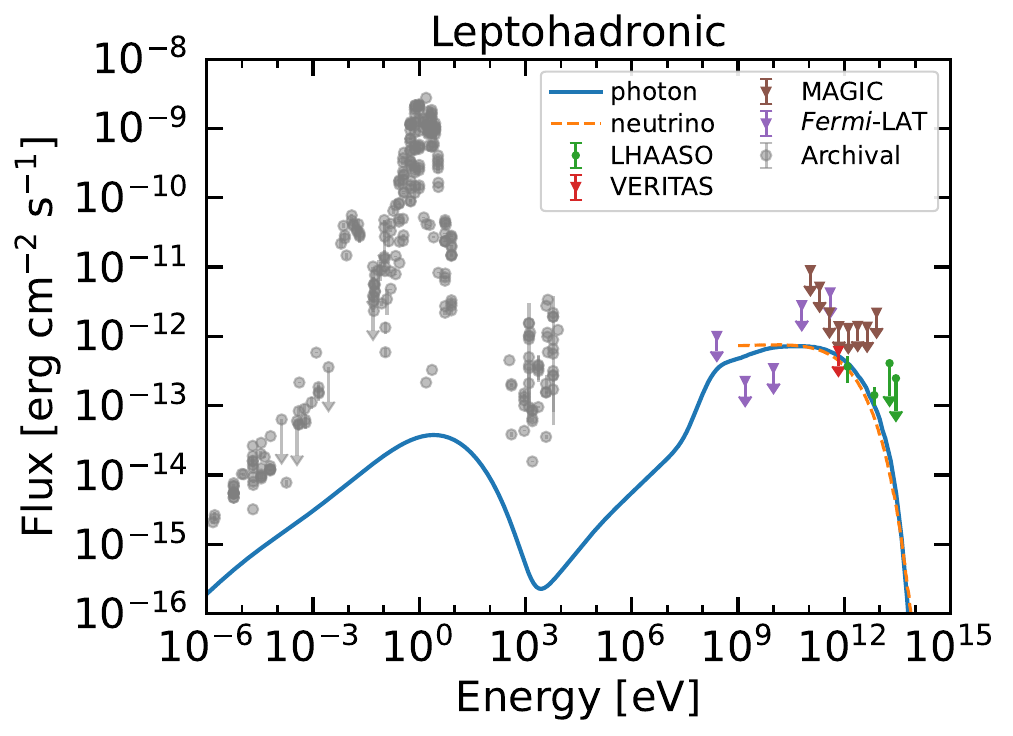}
    \end{minipage}
    \begin{minipage}{0.48\linewidth}
        \centering
        \includegraphics[width=\linewidth]{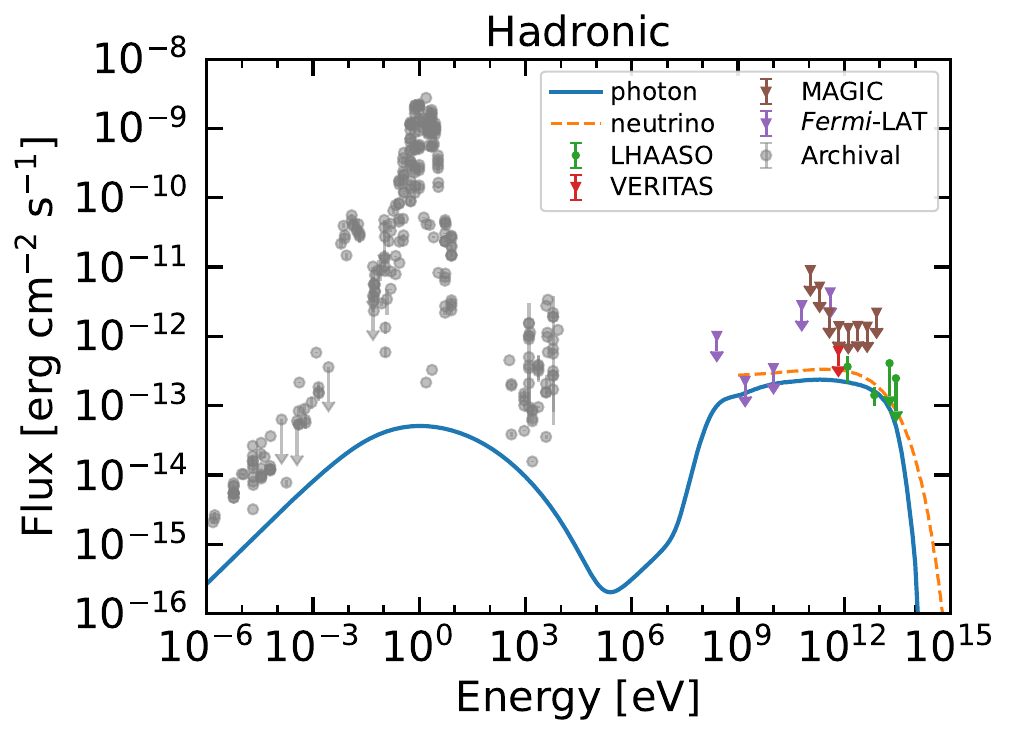}
    \end{minipage}
    \caption{
    \textit{Left}: Multiwavelength SEDs for the leptohadronic scenario.
    Adopted parameters are shown in Table~\ref{tab:parameters}.
    The blue curve shows the total nonthermal photon spectrum while the orange dashed line indicates the all-flavor neutrino spectrum.
    Green circles correspond to data from LHAASO during the quiet period \citep{2024ApJ...971L..45C}, while the red, purple, and brown triangles are 95\% confidence-level upper limits for the quiet period from VERITAS \citep{VERITAS_2026}, \textit{Fermi}-LAT \citep{VERITAS_2026}, and MAGIC \citep[post-flare dataset,][]{MAGIC_2026}, respectively.
    The gray thin symbols indicate archival data compiled by \citet{https://doi.org/10.26132/ned1}.
    \textit{Right}: Same as the left panel but for the hadronic scenario.
    }
\label{fig:MWLSEDs}
\end{figure*}

For IC scattering, we consider seed photons from stars of the host galaxy.
The archival multiwavelength SED of NGC~4278 from \citet{https://doi.org/10.26132/ned1} shows a bump at frequencies of $\sim10^{14}$--$10^{15}$~Hz with fluxes of $\sim10^{-9}$--$10^{-8}~\mathrm{erg~cm^{-2}~s^{-1}}$ (see Figure~\ref{fig:MWLSEDs}), whose corresponding luminosity ($\sim10^{44}~\mathrm{erg~s^{-1}}$) is significantly higher than the AGN luminosity \citep[$\sim10^{41}~\mathrm{erg~s^{-1}}$,][]{2010A&A...517A..33Y}.
We therefore interpret this emission as host-galaxy stellar radiation field. 
We model this photon field as a blackbody spectrum with a temperature of $0.4~\mathrm{eV}$ and the total luminosity of $L_\star=10^{44}~\mathrm{erg~s^{-1}}$, which reproduces archival infrared--optical data from \citet{https://doi.org/10.26132/ned1}.
We approximately calculate the photon energy density as $U_\mathrm{ph, \star}=3L_\star/(4\pi R_\star^2c)$, assuming a uniform, spherically symmetric emissivity within the central $R_\star\sim1$~kpc region.


\begin{figure*}
\centering
    \begin{minipage}{0.48\linewidth}
        \centering
        \includegraphics[width=\linewidth]{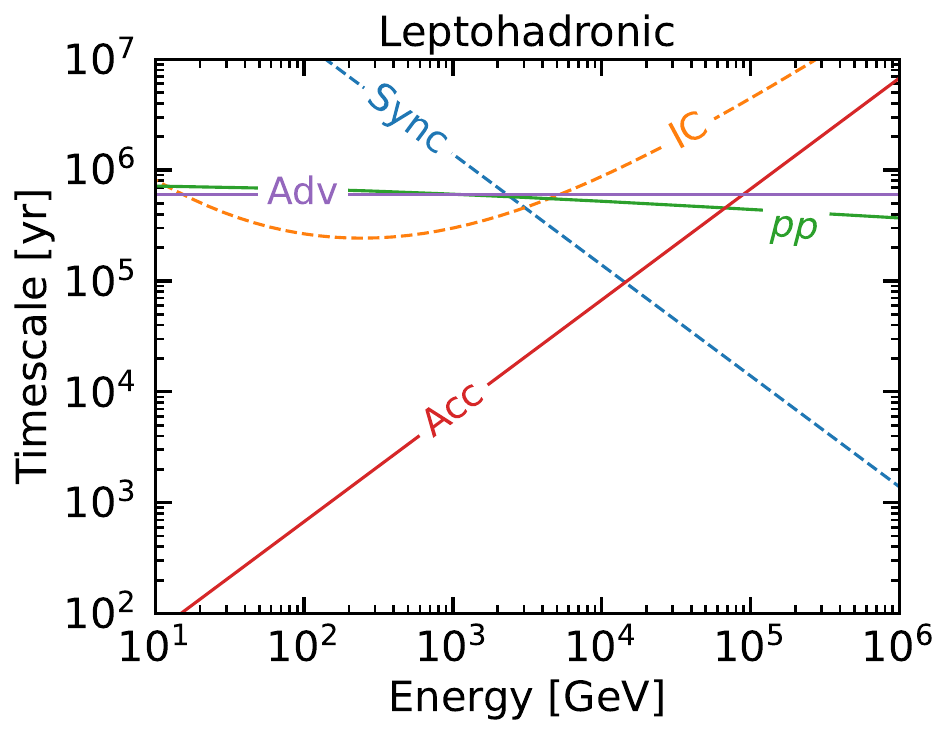}
    \end{minipage}
    \begin{minipage}{0.48\linewidth}
        \centering
        \includegraphics[width=\linewidth]{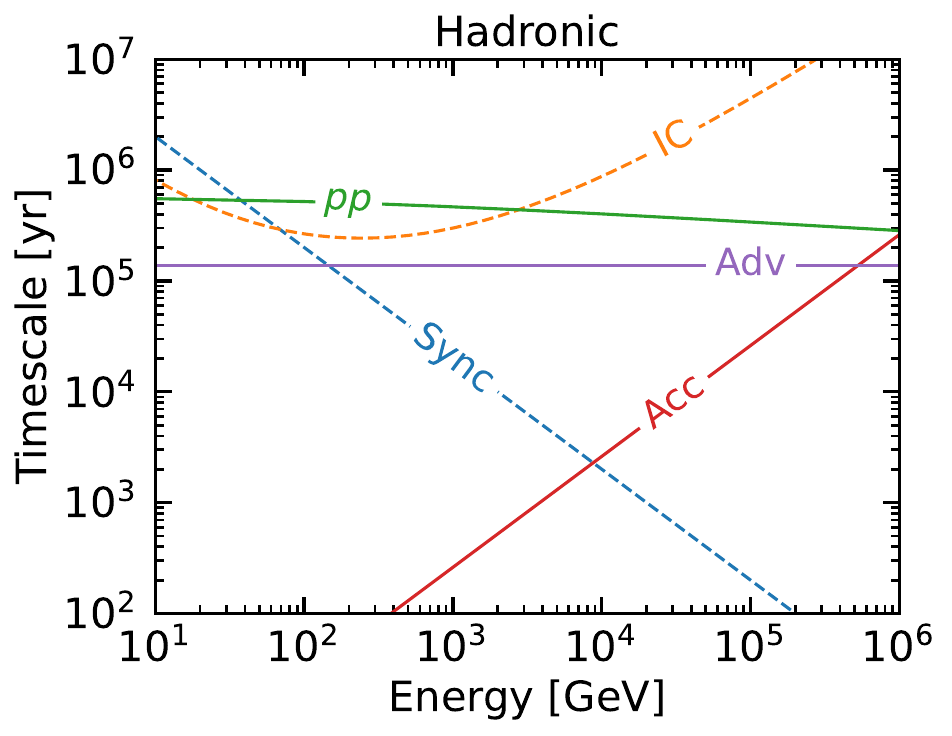}
    \end{minipage}
    \caption{
    \textit{Left}: Timescales of CR electrons and protons in the leptohadronic scenario.
    Solid lines indicate processes that occur for hadrons or equally for protons and electrons, while dashed lines show processes that occur only for electrons.
    Colors correspond to different processes, including synchrotron (blue), inverse Compton (IC, orange), \textit{pp} interactions (green), DSA (red), and advection (purple).
    \textit{Right}: Same as the left panel but for the hadronic scenario.
    }    
\label{fig:timescales}
\end{figure*}

\begin{table*}[t]
    \centering
    \begin{tabular}{lccc}
    \hline\hline
    Scenarios & NGC~4278 (leptohadronic) & NGC~4278 (hadronic) & Sgr~A* \\
    \hline
    \multicolumn{4}{c}{\textit{Varied values}}\\
    \hline 
    $\dot{K}_\mathrm{w}~(\mathrm{erg~s^{-1}})$ & $1.5\times10^{43}$ & $1.7\times10^{43}$ & $2.2\times10^{38}$ \\
    $R_\mathrm{FS}~(\mathrm{pc})$ & $2\times10^{2}$ & $8\times10^{1}$ & $1\times10^1$\\
    $B_\mathrm{ISM}~(\mu\mathrm{G})$ & 3 & $2.5\times10^1$ & $2.3\times10^2$ \\
    $q_\mathrm{CR}$ & 2.0 & 2.0 & 2.0 \\
    $n_0~(\mathrm{cm^{-3}})$ & $10^{2}$ & $1.3\times10^{2}$ & 64 \\
    $\epsilon_{e}/\epsilon_p$ & $10^{-1}$ & $10^{-2}$ & $10^{-3}$ \\
    \hline
    \multicolumn{4}{c}{\textit{Fixed values}} \\
    \hline
    $\epsilon_p$ & $10^{-1}$ & $10^{-1}$ & $10^{-1}$ \\
    $\xi_B$ & 1 & 1 & 1 \\
    \hline
    \multicolumn{4}{c}{\textit{Derived values}}\\
    \hline
    $V_\mathrm{FS}~(\mathrm{km~s^{-1}})$ & $3.2\times10^2$ & $5.7\times10^2$ & $6.8\times10^1$ \\
    $f_{pp}$ at 1~TeV & 1.00 & 0.30 & 0.15 \\
    $E_{p,\mathrm{cut}}~(\mathrm{TeV})$ & 48 & $4.2\times10^2$ & 64 \\
    \hline
    \end{tabular}
    \caption{
        \textit{Top}: Values of model parameters that we varied and adopted for NGC~4278 (the leptohadronic and hadronic scenarios) and Sgr~A*.
        \textit{Middle}: Values of model parameters that we fix for both NGC~4278 and Sgr~A*.
        \textit{Bottom}: Quantities derived from the varied parameters, shown for reference.
    }
    \label{tab:parameters}
\end{table*}

\section{Results}
\label{sec:results}

In this section, we aim to reproduce the TeV gamma-ray flux during the quiet period by the leptohadronic and hadronic scenarios.
Resulting timescales and SEDs are shown in Figures~\ref{fig:timescales} and \ref{fig:MWLSEDs}, respectively.

\subsection{Leptohadronic Scenario}\label{sec:leptohadronic}

In the leptohadronic scenario, the gamma-ray band is dominated by \textit{pp} interactions with a subdominant contribution from IC scattering, while \textit{pp} interactions dominate neutrino emission.
Left panels in Figures~\ref{fig:timescales} and \ref{fig:MWLSEDs} show timescales and SEDs, respectively.
The adopted parameters are $(\dot{K}_\mathrm{w}, R_\mathrm{FS}, B_\mathrm{ISM}, q_\mathrm{CR}, n_0, \epsilon_e/\epsilon_p)=(1.5\times10^{43}~\mathrm{erg~s^{-1}},\ 2\times10^2~\mathrm{pc},\ 3~\mu\mathrm{G},\ 2.0,\ 10^2~\mathrm{cm^{-3}},\ 10^{-1})$, summarized in Table~\ref{tab:parameters}.
The adopted radius is roughly comparable to the $\approx122$~pc outflow extent reported by \citet{2024A&A...683A..43H}.
The density is consistent with the ALMA-inferred value of $\sim10^2~\mathrm{cm^{-3}}$ \citep{2025PASJ..tmp..127S}.
We caution, however, that a uniform medium with $n_0=10^2~\mathrm{cm^{-3}}$ filling a sphere of $R_\mathrm{FS}=2\times10^2$~pc contains $\approx8\times10^7~M_\odot$, which is about eight times the molecular-gas mass of $\sim10^7~M_\odot$ measured on the 100-pc scale \citep{2025PASJ..tmp..127S}.
We therefore take $n_0$ to be the mean density of all gas phases swept up by the shock, and this benchmark implicitly assumes that atomic and ionized gas not traced by the CO observations (e.g., the warm ionized gas described in Section~\ref{sec:NGC4278 MWL}) contributes within $R_\mathrm{FS}$.
The spectral index $q_\mathrm{CR}=2.0$ is consistent with a standard DSA theory \citep[e.g.,][]{1983RPPh...46..973D}.
The adopted electron-to-proton ratio, $\epsilon_e/\epsilon_p=10^{-1}$, lies at the upper end of the range discussed in Section~\ref{sec:model emission}.
We therefore regard this scenario as an optimistic realization, rather than as a generic expectation.
As seen in the timescale figure, the wind system reaches the proton-calorimeter limit (i.e., $f_{pp}=1$) in the TeV band.
Cutoff energies of both electrons and protons are approximately a few tens of TeV.

This scenario is consistent with the sub-TeV and TeV gamma-ray upper limits, but in tension with the GeV ones.
The predicted flux at $680$~GeV is slightly below the VERITAS upper limit \citep{VERITAS_2026}, and it also stays below the MAGIC upper limits \citep{MAGIC_2026}.
However, the predicted flux at $\sim1$--10~GeV is a factor of $\approx2$ above the
95\% confidence-level \textit{Fermi}-LAT upper limits for the quiet period recently
reported by \citet{VERITAS_2026}.
We discuss in Section~\ref{sec:test} the implications for the choice between this scenario and the other scenario described in Section~\ref{sec:hadronic}.

In the GHz band ($\sim10^{-5}$~eV), primary electrons and secondary electrons from \textit{pp} interactions contribute about equally, and the predicted flux stays below the archival radio data.

\subsection{Hadronic Scenario}\label{sec:hadronic}

In the hadronic scenario, both gamma rays and high-energy neutrinos are dominated solely by \textit{pp} interactions.
The adopted parameters are $(\dot{K}_\mathrm{w}, R_\mathrm{FS}, B_\mathrm{ISM}, q_\mathrm{CR}, n_0, \epsilon_e/\epsilon_p)=(1.7\times10^{43}~\mathrm{erg~s^{-1}},\ 8\times10^1~\mathrm{pc},\ 25~\mu\mathrm{G},\ 2.0,\ 1.3\times10^2~\mathrm{cm^{-3}},\ 10^{-2})$, also summarized in Table~\ref{tab:parameters}.
Corresponding timescales and SEDs are shown in the right panels of Figures~\ref{fig:timescales} and \ref{fig:MWLSEDs}, respectively.
The biggest difference from the leptohadronic scenario is that the spectral shape around the LHAASO data is harder (see Figure~\ref{fig:HESEDs} for the comparison).
This is because the higher magnetic field lets the shock accelerate protons to higher energies (Equation~\ref{eq:E_p_max}).
The shock radius and gas density are also consistent with the MEGARA \citep{2024A&A...683A..43H} and ALMA \citep{2025PASJ..tmp..127S} observations, respectively.
In this case the swept-up mass, $\approx7\times10^6~M_\odot$, is comparable to the observed molecular-gas mass, so no additional gas reservoir needs to be invoked.
Cutoff energies of electrons and protons are around $\lesssim10$~TeV and $>10^2$~TeV, respectively.
Unlike the leptohadronic scenario, the wind is not a proton calorimeter here (see the right panel of Figure~\ref{fig:timescales}).
The predicted flux at $680$~GeV is below the VERITAS \citep{VERITAS_2026} and MAGIC \citep{MAGIC_2026} upper limits, and by contrast to the leptohadronic scenario, this scenario is consistent also with the \textit{Fermi}-LAT upper limits \citep{VERITAS_2026}.

The predicted radio flux comes closer to the archival data around $\sim10^{-6}$--$10^{-3}$~eV than in the leptohadronic case because of the higher magnetic field, and it is dominated ($\approx80\%$) by secondary electrons from \textit{pp} interactions, since $\epsilon_e/\epsilon_p$ is ten times smaller here.


\section{Discussion}
\label{sec:discussion}

In this section, we discuss whether the required wind powers for the TeV emission in NGC~4278 are reasonable or not, how to observationally test the wind-emission scenarios, and how much an LLAGN population contributes to the observed extragalactic neutrino background.

\subsection{Gas Accretion Scenarios and Wind Powers}
\label{sec:gas accretion}

We investigate the validity of the wind powers required in the leptohadronic and hadronic scenarios.
In the framework of the adiabatic inflow-outflow solutions \citep[ADIOS,][]{1999MNRAS.303L...1B}, the energy budget of an outflow is ultimately supplied by the power of gas accretion. 
We first estimate $\dot{K}_\mathrm{w}$. 
Then, we consider three accretion scenarios, involving the warm ionized gas, hot X-ray gas, and cold molecular gas introduced in Section~\ref{sec:NGC4278 MWL}. 

First, we estimate $\dot{K}_\mathrm{w}$ for an RIAF wind following ADIOS.
The central SMBH accretes ambient gas with a number density of $n_\infty$, a temperature of $T_\infty$, and an adiabatic sound speed of $c_{\mathrm{s},\infty}=\sqrt{\gamma_{\mathrm{gas}} k_{\mathrm{B}} T_\infty/(\mu_\infty m_p)}$, where $\gamma_{\mathrm{gas}}=5/3$ is the adiabatic index and $\mu_\infty$ is the mean molecular weight of the accreted gas. 
The accretion profile at a radius of $R$ normalized at the Bondi radius $R_\mathrm{B}=GM_\mathrm{BH}/c_{\mathrm{s},\infty}^2$ \citep{1952MNRAS.112..195B} is
\begin{align}\label{eq:accretion profile ADIOS}
    \dot{M}_\mathrm{acc}(R)
    =\alpha_\mathrm{v}\,\dot{M}_\mathrm{B}\left(\frac{R}{R_\mathrm{B}}\right)^s,
\end{align}
where $s$ is the radial index of the accretion-rate profile, $\alpha_\mathrm{v}$ is the viscosity parameter, and $\dot{M}_\mathrm{B}=4\pi R_\mathrm{B}^2 \mu_\infty m_p n_\infty c_{\mathrm{s},\infty}$ is the Bondi accretion rate.
However, this would overestimate the real accretion rate when the distribution scale of gas $R_\mathrm{gas}$, which can feed the SMBH, is much smaller than the Bondi radius.
Practically, multiplying Equation~(\ref{eq:accretion profile ADIOS}) by $(R_\mathrm{gas}/R_\mathrm{B})^2$ would make it more accurate:
\begin{align}\label{eq:accretion profile}
    \dot{M}_\mathrm{acc}(R)
    =\alpha_\mathrm{v}\,\dot{M}_\mathrm{B}\left(\frac{R}{R_\mathrm{B}}\right)^s\min\left[1, (R_\mathrm{gas}/R_\mathrm{B})^{2}\right].
\end{align}
The differential mass-outflow rate between $R$ and $R+dR$ can be estimated as 
\begin{align}
    d\dot{M}_\mathrm{w}(R)=\dot{M}_\mathrm{acc}(R+dR)-\dot{M}_\mathrm{acc}(R)=\frac{d\dot{M}_\mathrm{acc}}{dR}dR,
\end{align}
where $\dot{M}_\mathrm{acc}(R)$ is adopted from Equation~(\ref{eq:accretion profile}).
We take the escape velocity at each radius as the wind velocity: $v_\mathrm{w}(R)=\sqrt{2GM_\mathrm{BH}/R}=(R/R_\mathrm{Sch})^{-1/2}c$, where $R_\mathrm{Sch}=2GM_\mathrm{BH}/c^2$ is the Schwarzschild radius.
Then, the cumulative kinetic power of the RIAF-driven wind from the wind-launching inner radius $R_\mathrm{l}$ to $R_\mathrm{B}$ becomes
\begin{align}
    \dot{K}_\mathrm{w}&=\int_{R_\mathrm{l}}^{R_\mathrm{B}}\frac{1}{2}d\dot{M}_\mathrm{w}(R)v_\mathrm{w}^2(R)\nonumber\\
    &=\frac{s}{2(1-s)}\dot{M}_\mathrm{acc}(R_\mathrm{l})v_\mathrm{w}^2(R_\mathrm{l})\left[1-\left(\frac{R_\mathrm{l}}{R_\mathrm{B}}\right)^{1-s}\right]\label{eq:power}.
\end{align}
In this study, we fix $R_\mathrm{l}=5R_\mathrm{Sch}$ \citep{2014ARA&A..52..529Y} and $\alpha_\mathrm{v}=0.1$ \citep[e.g.,][]{2008bhad.book.....K}, while we consider the shallower accretion profile of $s=0.3$ inferred from observations \citep[e.g.,][]{2003ApJ...598..301Y} and the steeper profile of $s=0.5$ from numerical simulations \citep[e.g.,][]{2012ApJ...761..129Y}.

We find that in NGC~4278, the warm ionized gas can supply sufficient power to the wind for the shallower accretion profile, but not for the steeper one.
As described in Section~\ref{sec:NGC4278 MWL}, \citet{2024A&A...683A..43H} reported ionized gas extending to $R_\mathrm{gas}\sim100$~pc.
We assume that a part of the ionized gas phase can be an effective feeding reservoir appearing as an inflow \citep{2018MNRAS.480.1106C}.
We adopt an electron number density of $300~\mathrm{cm^{-3}}$ \citep{2024A&A...683A..43H}, $T_\infty=10^4$~K, $\mu_\infty=1.4$, and $s=0.3$ for the shallower profile. 
We caution, however, that the adopted density \citep[$300~\mathrm{cm^{-3}}$,][]{2024A&A...683A..43H} is measured for the outflowing gas and may not represent the ambient gas available for accretion.
Using these values, the estimated accretion power at $R=5R_\mathrm{Sch}$ and wind power from Equations~(\ref{eq:accretion profile}) and (\ref{eq:power}) become $\dot{M}_\mathrm{acc}c^2\sim3\times10^{44}~\mathrm{erg~s^{-1}}$ and $\dot{K}_\mathrm{w}\sim10^{43}~\mathrm{erg~s^{-1}}$, respectively.
This wind power is comparable to those required in both scenarios.
In the steeper case (i.e., $s=0.5$), the estimated power becomes $\dot{K}_\mathrm{w}\lesssim10^{42}~\mathrm{erg~s^{-1}}$, which is smaller than the required power.
This is because in the steeper case, more gas is lost as outflow at the outer regions, where the ejection (i.e., escape) velocity is slower.
Therefore, determining $s$ values is crucial for wind dynamics.

We compare our accretion model with the accretion rate independently inferred from X-ray observations.
\citet{Das_2026} recently reported NuSTAR hard X-ray observations of NGC~4278 and, fitting the quiet-period data with an RIAF model, inferred an accretion power of $\dot{M}_\mathrm{acc}c^2\approx1.1\times10^{-3}L_\mathrm{Edd}\approx4\times10^{43}~\mathrm{erg~s^{-1}}$ at $R=10R_\mathrm{Sch}$.
This is about an order of magnitude lower than the inner accretion power predicted by our warm-ionized-gas model with $s=0.3$.
We do not regard this tension as fatal, because the X-ray-inferred value depends on the adopted RIAF prescription and does not account for the mass carried away by a wind.
A self-consistent RIAF model including winds would be needed for a direct comparison.

In contrast, it is difficult for accretion of the hot X-ray gas and the molecular gas to supply sufficient power for the TeV emission even in the shallower accretion profile.
We first estimate the wind power supplied by Bondi accretion of the hot gas as $\dot{K}_\mathrm{w}\approx8\times10^{39}~\mathrm{erg~s^{-1}}$, which is much lower than the values required by our TeV modeling, adopting $n_\infty=0.1~\mathrm{cm^{-3}}$, $T_\infty=0.6$~keV, $\mu_\infty=0.6$ (see Section~\ref{sec:NGC4278 MWL} for detail), $s=0.3$ \citep{2003ApJ...598..301Y}, and $R_\mathrm{gas}\sim4~\mathrm{kpc}$ \citep[$\gg R_\mathrm{B}$,][]{2012ApJ...758...65B, 2012ApJ...758...94P}.
For the molecular gas, adopting $n_\infty=100~\mathrm{cm^{-3}}$ \citep{2025PASJ..tmp..127S}, $T_\infty=10^2~\mathrm{K}$ \citep{2011MNRAS.414.1827T}, $\mu_\infty=2.3$, $s=0.3$ \citep{2003ApJ...598..301Y}, and $R_\mathrm{gas}\sim10^2$~pc \citep{2025PASJ..tmp..127S} leads to $\dot{K}_\mathrm{w}\sim10^{41}~\mathrm{erg~s^{-1}}$, which is also smaller than the required power.
For the steeper profile with $s=0.5$, the estimated kinetic powers for both gas phases are well below the required ones in both the leptohadronic and hadronic scenarios.

\subsection{Model Discrimination and Future Tests}\label{sec:test}

\begin{figure}
    \centering
    \includegraphics[width=\linewidth]{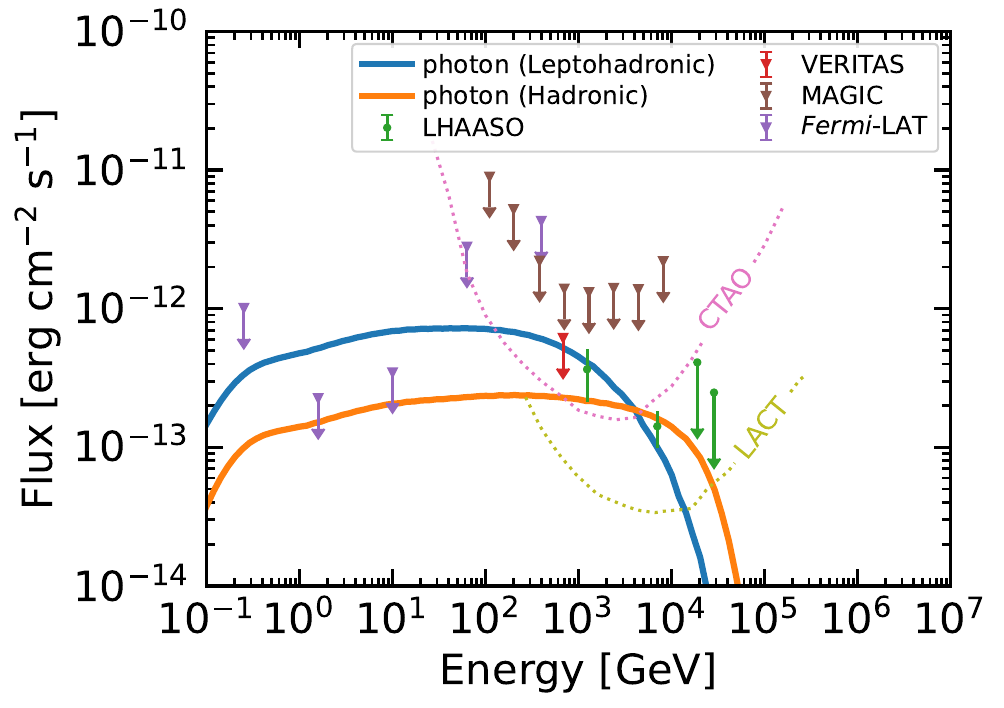}
    \caption{
    Gamma-ray SEDs for the leptohadronic (blue) and hadronic (orange) scenarios.
    Green circles show LHAASO data obtained during the quiet period \citep{2024ApJ...971L..45C}, while the red, purple, and brown triangles are 95\% confidence-level upper limits for the quiet period from VERITAS \citep{VERITAS_2026}, \textit{Fermi}-LAT \citep{VERITAS_2026}, and MAGIC \citep[post-flare dataset,][]{MAGIC_2026}, respectively.
    The pink and olive dotted lines indicate the sensitivities of the CTAO \citep[$\sim50$~hr,][]{2019scta.book.....C} and the LACT \citep[$\sim500$~hr,][]{2025ChPhC..49c5001Z}.
    }
    \label{fig:HESEDs}
\end{figure}

In this section, we discuss how to discriminate between the leptohadronic and hadronic
scenarios and how to test them with future observations.
The current multiwavelength data already prefer the hadronic scenario over the
leptohadronic one.
As shown in Section~\ref{sec:leptohadronic}, the leptohadronic benchmark exceeds the
quiet-period \textit{Fermi}-LAT upper limits of \citet{VERITAS_2026} by a factor of
$\approx2$ at $\sim1$--10~GeV, whereas the hadronic benchmark lies below every reported
limit.
If these limits hold, they indicate a limitation of the leptohadronic benchmark.
We caution, however, that \textit{Fermi}-LAT analyses of NGC~4278 have been mutually
inconsistent (Section~\ref{sec:NGC4278 gamma}), and this single set of limits should not
be weighted more heavily than the others.
With this caveat in mind, we regard the hadronic scenario as favored at present, although
the leptohadronic one is not strongly excluded.
That existing observations can already discriminate between two wind models demonstrates that the wind-emission scenario is testable, and implies that
future observations will be able to test our scenarios from a more multifaceted perspective.

We now turn to future radio, gamma-ray, and neutrino observations.
First, imaging wind emission is crucial for distinguishing among several scenarios.
If radio observations resolve the wind-like quasi-spherical morphology and the detected fluxes are comparable to either expectation (the leptohadronic and/or hadronic scenario), that would support the wind scenarios.
A VLBI observation was previously conducted for a small-scale region, $\lesssim100$~mas \citep{2005ApJ...622..178G}, which did not construct a radio image of the predicted shock scale (i.e., 100~pc scale).
To resolve the wind morphology in our scenarios, angular resolutions comparable to the angular diameter $\theta_\mathrm{s}\approx2R_\mathrm{FS}/D\approx2.1''$ (hadronic) to $5.2''$ (leptohadronic) are required.
The VLA can reach these resolutions\footnote{\url{https://science.nrao.edu/facilities/vla/docs/manuals/oss/performance/resolution}}.
The VLA can also detect the predicted radio fluxes in our scenarios.
Our predictions at 1~GHz are $\approx37$~mJy and $\approx55$~mJy for the leptohadronic and hadronic scenarios, respectively.
Modeling the radio emission as a uniform disk of $\theta_\mathrm{s}$, we obtain a source
solid angle of $\Omega_\mathrm{s}=\pi\theta_\mathrm{s}^2/4\approx21~\mathrm{arcsec^2}$
($3.3~\mathrm{arcsec^2}$) for the leptohadronic (hadronic) scenario.
For a circular Gaussian restoring beam of FWHM $\theta_\mathrm{b}$, the beam solid
angle is $\Omega_\mathrm{b}=\pi\theta_\mathrm{b}^2/(4\ln2)\approx1.13\,\theta_\mathrm{b}^2$
\citep{2016era..book.....C}, so that the flux densities quoted above correspond to
surface brightnesses of $\approx2$ and $\approx19~\mathrm{mJy~beam^{-1}}$ for
$\theta_\mathrm{b}=1''$.
These surface brightnesses are well above the VLA sensitivity\footnote{\url{https://science.nrao.edu/facilities/vla/docs/manuals/oss/performance/sensitivity}}.
Therefore, the VLA could distinguish not only between the leptohadronic and hadronic scenarios but also between the wind scenario and other compact scenarios.

Future telescopes should be able to detect TeV gamma rays from NGC~4278 in both scenarios.
Figure~\ref{fig:HESEDs} shows TeV gamma-ray SEDs for the scenarios together with the sensitivities of the Cherenkov Telescope Array Observatory \citep[CTAO, $\sim50$~hr,][]{2019scta.book.....C} and the Large Array of Imaging Atmospheric Cherenkov Telescopes \citep[LACT, $\sim500$~hr,][]{2025ChPhC..49c5001Z} at the LHAASO site.
In the figure, we can see that the predicted fluxes by both scenarios exceed the sensitivities of both CTAO and LACT.
Therefore, gamma-ray observations will be crucial for testing the wind-emission scenarios.

Those telescopes could also test our wind scenario by measuring the high-energy cutoff, corresponding to the maximum CR proton energy.
Our two scenarios predict markedly different cutoffs, $E_{p, \mathrm{cut}}\approx48$~TeV for the leptohadronic case and $\approx4.2\times10^2$~TeV for the hadronic one, which can be distinguished with CTAO and LACT.
These predictions rest on the Bohm factor $\xi_B$, which we have fixed to unity, so we first show that it can be at most a few.
Equations~(\ref{eq:shock velocity}), (\ref{eq:DSA timescale}), and (\ref{eq:advection timescale}) give
\begin{align}\label{eq:E_p_max}
    E_{p, \mathrm{cut}}\propto \xi_B^{-1}B_\mathrm{ISM}R_\mathrm{FS}^{1/3}(\dot{K}_\mathrm{w}/n_0)^{1/3},
\end{align}
which shows that a larger $\xi_B$ must be offset by a larger $B_\mathrm{ISM}$, $R_\mathrm{FS}$, or $\dot{K}_\mathrm{w}$, or by a smaller $n_0$, to keep reproducing the LHAASO data.
The power $\dot{K}_\mathrm{w}$ can be raised by at most a factor of $\sim2$, because the required wind power is already a few tens of percent of the accretion power inferred by \citet[][see Section~\ref{sec:gas accretion}]{Das_2026}.
Consequently, the density $n_0$ cannot be much lower, since the gamma-ray flux is proportional to $\epsilon_p\dot{K}_\mathrm{w}f_{pp}\propto \dot{K}_\mathrm{w}^{2/3}n_0^{4/3}R_\mathrm{FS}^{5/3}$.
A larger $R_\mathrm{FS}$ does not help either because it raises the synchrotron flux, which scales as $B_\mathrm{ISM}^{(q_\mathrm{CR}+1)/2}R_\mathrm{FS}^{5/3}$ \citep[][]{1986rpa..book.....R} when radio-emitting electrons advect faster than they cool as in our scenarios, far more steeply than it raises $E_{p,\mathrm{cut}}$.
The magnetic field therefore has to carry the entire compensation and our scenarios lie at most an order of magnitude below the archival radio fluxes, which permits $B_\mathrm{ISM}$ to increase by at most a factor of $\approx10^{2/(q_\mathrm{CR}+1)}\sim5$ for $q_\mathrm{CR}=2$ and hence $\xi_B\lesssim5$.
A wind scenario with $\xi_B\gg5$ \citep[e.g.,][for AGN jets]{2016ApJ...828...13I} would overshoot the radio data.
The predicted cutoff energies are thus conservative upper bounds.
Since $\xi_B\geq1$ by definition and $B_\mathrm{ISM}$ can rise by at most $\sim5$, $E_{p,\mathrm{cut}}$ cannot exceed $\sim2$~PeV. 
This implies that a gamma-ray cutoff energy within our wind framework is at most $\sim200$~TeV because gamma rays from \textit{pp} interactions carry $\sim10\%$ of the parent proton energy \citep[e.g.,][]{2006PhRvD..74c4018K}.
We stress, however, that such a high cutoff cannot be measured directly, because EBL and CMB attenuation makes NGC~4278 opaque above $\approx36$~TeV, so that a cutoff inferred from the observed spectrum above this energy would reflect propagation rather than the source.
The practical discriminator is instead the spectral shape below a few tens of TeV: even after attenuation, the two scenarios differ by a factor of $\sim2$ at 10~TeV and $\sim10$ at 30~TeV.
Together with radio observations that pin down the magnetic field, TeV gamma-ray observations can thus constrain the maximum energy and provide strong tests of our wind scenarios.

MeV observations could also be crucial for constraining models.
For example, the Compton Spectrometer and Imager \citep[COSI,][]{2024icrc.confE.745T} will have a sensitivity of $\gtrsim 10^{-11}~\mathrm{erg~cm^{-2}~s^{-1}}$ at $\sim1$~MeV. 
COSI can place upper limits around MeV energies for NGC~4278 and could rule out some models \citep[e.g.,][which predicts $\sim 10^{-11}~\mathrm{erg~cm^{-2}~s^{-1}}$ at $\sim1$~MeV using a jet model]{2024ApJS..271...10W}.

\begin{figure}
    \centering
    \includegraphics[width=\linewidth]{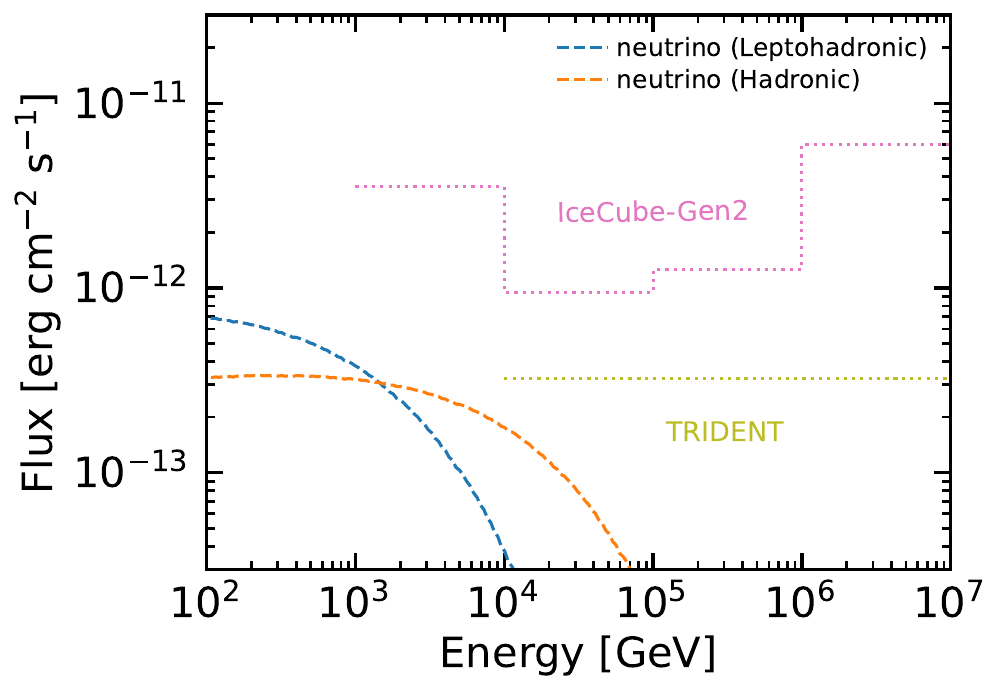}
    \caption{
    Similar to Figure~\ref{fig:HESEDs} but for all-flavor neutrinos.
    The blue and orange dashed curves correspond to the leptohadronic and hadronic scenarios.
    The pink dotted steps and the olive dotted line indicate the 90\%~confidence-level
    sensitivities for 10~yr of IceCube-Gen2 \citep[][]{2021JPhG...48f0501A} and TRIDENT
    \citep[][]{2023NatAs...7.1497Y} observations, respectively.
    Note that the former is quasi-differential, assuming a spectral index of 2.0 within each
    energy bin, whereas the latter is integrated, assuming the same index over the whole energy range.
    We multiply the original sensitivities in the literature by a factor of 3 to convert from per-flavor fluxes to all-flavor ones, assuming a flavor ratio of 1:1:1.
    }
    \label{fig:HEnuSEDs}
\end{figure}

On the other hand, it is difficult for future telescopes to detect TeV--PeV neutrinos from NGC~4278.
Figure~\ref{fig:HEnuSEDs} shows the neutrino fluxes (all-flavor and $\nu+\bar{\nu}$) predicted in the leptohadronic and hadronic scenarios together with the flux sensitivities of IceCube-Gen2 \citep{2021JPhG...48f0501A} and Tropical Deep-sea Neutrino Telescope \citep[TRIDENT,][]{2023NatAs...7.1497Y}, both future TeV--PeV neutrino telescopes.
The neutrino fluxes at 10~TeV in both scenarios are lower than the TRIDENT and IceCube-Gen2 sensitivities, implying that they are unlikely to be able to detect NGC~4278.
If a brighter TeV state recurs frequently or persists for a sufficiently long duration, the corresponding neutrino flux may approach the sensitivity of future detectors.

\subsection{Contribution of an RIAF Wind Population to the Extragalactic Neutrino Background}\label{sec:DNB}

We estimate how much an LLAGN population contributes to the observed extragalactic neutrino background, assuming an NGC~4278 environment for all sources.
\citet{2026ApJ..1003...71Y} suggest that a leptohadronic wind scenario can account for the PeV extragalactic neutrino background via photomeson interactions, but we examine that scenario considering \textit{pp} interactions. 
The number density of LLAGNs is $\sim10^{-3}~\mathrm{Mpc^{-3}}$ \citep{2008ARA&A..46..475H} around a H$\alpha$ luminosity of $\approx4\times10^{39}~\mathrm{erg~s^{-1}}$, which is comparable to that of NGC~4278 \citep{2001ApJ...549L..51H}.
We consider no evolution of the LLAGN population in number density, which gives a redshift evolution factor of $\xi_z\approx0.6$ \citep{1998PhRvD..59b3002W}.
We assume all LLAGNs have NGC~4278-like parameters, including a CR luminosity of $1.5\times10^{42}~\mathrm{erg~s^{-1}}$, \textit{pp} interaction efficiency $f_{pp}\approx0.3$, and the maximum proton energy $E_{p, \mathrm{cut}}\approx4\times10^{2}~\mathrm{TeV}$, all of which are those of the hadronic scenario (Section~\ref{sec:hadronic}).

We find that in the RIAF-driven wind scenario, NGC~4278 is unlikely to be a representative LLAGN in the neutrino sky. 
Adopting the above assumptions and NGC~4278-like parameters, Equation~(7) in \citet{2013PhRvD..88l1301M} yields $\approx8\times10^{-8}~\mathrm{GeV~cm^{-2}~s^{-1}~sr^{-1}}$ below 20~TeV. 
This exceeds the IceCube observations $\sim5\times10^{-8}~\mathrm{GeV~cm^{-2}~s^{-1}~sr^{-1}}$ at the same energy \citep{2026PhRvL.136l1002A}, implying that NGC~4278 is a powerful proton accelerator and/or has a high \textit{pp} efficiency, compared with a typical LLAGN.
More generally, the neutrino spectral index of $\approx2.8$ measured by IceCube \citep{2026PhRvL.136l1002A} rules out a scenario where \textit{pp}-induced neutrinos dominate the TeV--PeV neutrino sky \citep[see][for details]{2013PhRvD..88l1301M,Murase2016}.
However, if RIAF winds with lower CR power and \textit{pp} efficiency are ubiquitous among LLAGNs, they may still contribute subdominantly to the extragalactic neutrino background. 
Alternatively, RIAF disks themselves may give a larger contribution to neutrinos both in the direction to NGC~4278 \citep{Das_2026} and in the extragalactic neutrino background \citep{Kimura2015,Kimura2021}.  

\section{Application to Sgr~A*}\label{sec:Sgr A*}

To generalize the importance of an RIAF wind, we investigate whether it can account for TeV emission from Sgr~A*. 
It is an LLAGN at the Galactic center from which TeV gamma rays have been detected.
The distance from Earth and SMBH mass are $D\approx8.2$~kpc \citep{2019A&A...625L..10G} and $\approx4\times10^6~M_\odot$ \citep{2016ApJ...830...17B}.
The current bolometric luminosity is $\sim10^{36}~\mathrm{erg~s^{-1}}$ \citep[e.g.,][]{2003ApJ...598..301Y}, corresponding to $\sim2.5\times10^{-9}L_\mathrm{Edd}$.
Observations have revealed multiple gas phases around Sgr~A*, including molecular gas and hot ionized gas observed in the X-ray band.
Radio observations have revealed the central molecular zone (CMZ) extending to the central $\sim100$~pc-scale with a typical density of $\sim10^2$--$10^3~\mathrm{cm^{-3}}$ \citep[e.g.,][]{2025ApJ...984..156B}.
X-ray observations have revealed hot gas within $\sim1$~pc \citep{2003ApJ...591..891B} with a density of $\sim10^2~\mathrm{cm^{-3}}$ \citep[e.g.,][]{2003ApJ...598..301Y}.
Radio observations have confirmed nonthermal emission from the central $\sim10$~pc region \citep[e.g.,][]{2016ApJ...817..171Z, 2016ApJ...819...60Y, 2020MNRAS.499.3909Y, 2022ApJ...925..165H}, which could originate from a shock driven by an RIAF wind.
In the TeV band, the High Energy Stereoscopic System (H.E.S.S.) detected Sgr~A* \citep[HESS~J1745$-$290, e.g.,][]{2016Natur.531..476H}.
They separate the gamma-ray emission into a diffuse component and a compact component within 10~pc, suggesting that a particle accelerator within the compact regions may be connected to the SMBH.

Theoretical works suggest that the TeV emission around Sgr~A* originates from an inner RIAF activity.
\citet{2015PhRvD..92b3001F} found that their model can explain the TeV emission if the accretion rate on Sgr~A* was much larger in the past than it is today. 
\citet{2017JCAP...04..037F} further suggest that Sgr~A* injected a large number of PeV CRs at an outburst about $10^7~\mathrm{yr}$ ago.
Given that these studies support an RIAF scenario for the TeV emission, we hypothesize that the RIAF-related activity can account for the emission.
We note that while the above studies consider CR acceleration in an RIAF itself, we consider CRs that are accelerated at shocks induced by RIAF-driven winds.
Since the shock-acceleration region is a point-like source for H.E.S.S., our one-zone model of an RIAF wind can be applied.

\begin{figure}
    \centering
    \includegraphics[width=\linewidth]{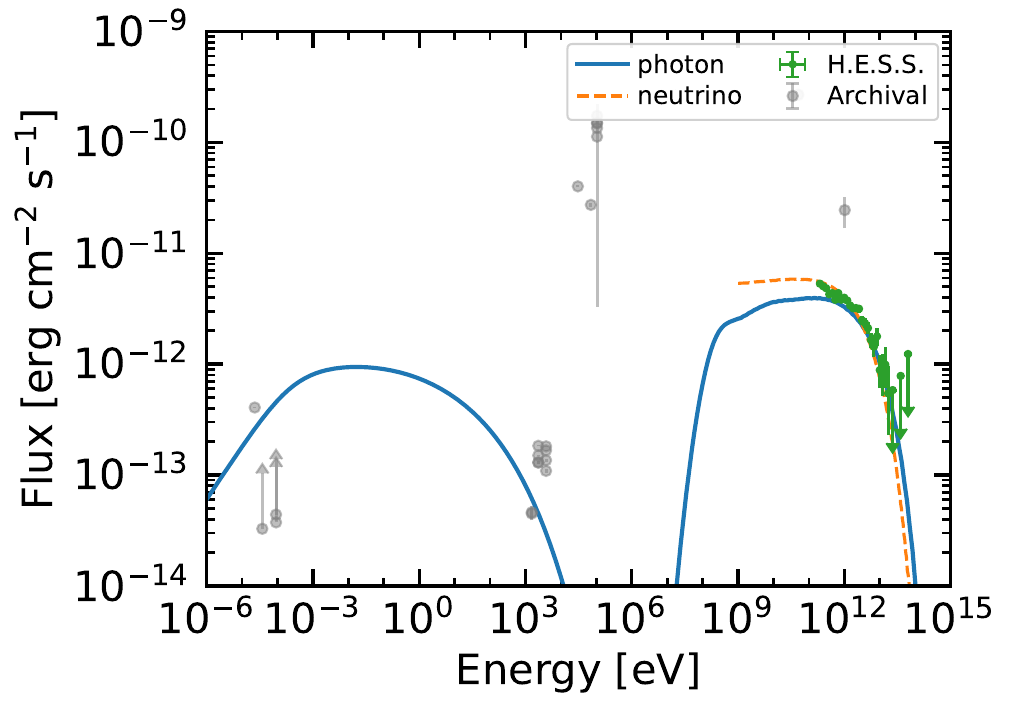}
    \caption{
    Similar to Figure~\ref{fig:MWLSEDs} but for Sgr~A*, with green points indicating gamma-ray data from H.E.S.S. \citep{2016Natur.531..476H}.
    Among the archival data reported by \citet{https://doi.org/10.26132/ned1}, the two $22$~GHz VERA points are the only ones we convert into lower limits, because their interferometric resolution is much finer than the shock scale (see Section~\ref{sec:Sgr A*}).
    }
    \label{fig:MWSED SgrA*}
\end{figure}

To reproduce the TeV flux, we treat model parameters similarly to the NGC~4278 case. 
We vary $\dot{K}_\mathrm{w}$, $R_\mathrm{FS}$, $B_\mathrm{ISM}$, $q_\mathrm{CR}$, $n_0$, and $\epsilon_e/\epsilon_p$.
We do not consider IC scattering as the photon energy density around the $\sim10$~pc region is much lower than the magnetic one due to the very weak AGN activity.
The values of the other parameters (i.e., $\epsilon_p$ and $\xi_B$) are fixed to the same values as in the NGC~4278 case (see Table~\ref{tab:parameters}). 

With reasonable parameters, our model reproduces the TeV gamma-ray flux.
Figure~\ref{fig:MWSED SgrA*} shows the resulting SEDs.
Here, we choose $(\dot{K}_\mathrm{w}, R_\mathrm{FS}, B_\mathrm{ISM}, q_\mathrm{CR}, n_0, \epsilon_e/\epsilon_p)=(2.2\times10^{38}~\mathrm{erg~s^{-1}},\ 10~\mathrm{pc},\ 2.3\times10^{2}~\mu\mathrm{G},\ 2.0,\ 64~\mathrm{cm^{-3}},\ 10^{-3})$.
Table~\ref{tab:parameters} summarizes these adopted values and other fixed parameters.
In the Sgr~A* case the wind is far from the proton-calorimeter limit, with $f_{pp}\approx0.15$. 
Since dense molecular clouds occupy only a small fraction of the CMZ volume \citep[filling factor $< 0.1$, e.g.,][]{2019ApJ...883...54O}, the forward shock propagates mainly through the diffuse inter-cloud medium, whose volume-averaged density is well below the cloud density of $\sim10^2$--$10^3~\mathrm{cm^{-3}}$. 
Our adopted $n_0=64~\mathrm{cm^{-3}}$ should thus be interpreted as this volume-averaged density. 
The adopted forward-shock radius is consistent with radio observations \citep[e.g.,][]{2016ApJ...817..171Z, 2016ApJ...819...60Y, 2020MNRAS.499.3909Y, 2022ApJ...925..165H} and gamma-ray observations \citep{2016Natur.531..476H}.
To avoid overshooting archival radio data from \citet{https://doi.org/10.26132/ned1}, we adopt $\epsilon_e/\epsilon_p=10^{-3}$, which does not affect the high-energy emission as hadronic \textit{pp} interactions dominate the gamma-ray emission.
We treat the two $22$~GHz VERA points \citep{2007AJ....133.2487P} as lower limits, although \citet{https://doi.org/10.26132/ned1} does not, because their interferometric resolution may resolve out the extended emission.

We can also consider a reverse-shock interpretation.
High-spatial-resolution radio observations also revealed fast outflows from Sgr~A* on $\lesssim0.1$~pc scale \citep{2019ApJ...872....2R, 2020MNRAS.499.3909Y}.
If we interpret those as tracers of the reverse shock, the CR acceleration site can also be $\lesssim0.1$~pc.
This radius is comparable to that predicted from the balance between the wind ram pressure and pressure by \textit{Chandra}-detected gas with a temperature of $\sim1$~keV and a density of $\sim10^2~\mathrm{cm^{-3}}$ \citep{2003ApJ...598..301Y}.
After acceleration, CR protons may escape from the reverse-shocked region and interact with hydrogen nuclei in the CMZ.
We note that this is only a qualitative alternative, which the present data cannot exclude.
So far, both the forward-shock and reverse-shock interpretations remain possible.

Our results indicate that the TeV emission in Sgr~A* can originate from an RIAF wind powered by accretion of the X-ray-emitting hot gas.
As in Section~\ref{sec:gas accretion}, we estimate wind powers for the accretion of the observed gas in the X-ray band.
We adopt $s=0.3$ based on the observational work for Sgr~A* by \citet{2003ApJ...598..301Y}. 
Adopting $(n_\infty, \mu_\infty, T_\infty)=(10^2~\mathrm{cm^{-3}}, 0.6, 1~\mathrm{keV})$ for X-ray gas \citep{2003ApJ...598..301Y} and using Equation~(\ref{eq:power}), we derive the wind power as $\dot{K}_\mathrm{w}\sim4\times10^{38}~\mathrm{erg~s^{-1}}$, which is sufficient to supply the required $2.2\times10^{38}~\mathrm{erg~s^{-1}}$.
We note that this value is roughly consistent with that estimated in the previous work on an RIAF wind in Sgr~A* \citep[$\dot{K}_\mathrm{w}\sim1.5\times10^{38}~\mathrm{erg~s^{-1}}$,][]{2006PASJ...58..965T}.
The required spectral index of injected CRs $q_\mathrm{CR}=2.0$ is also consistent with a standard DSA theory \citep[e.g.,][]{1983RPPh...46..973D}.


\section{Conclusions}
\label{sec:conclusion}

In summary, using \textsc{AMES}, we investigated RIAF-driven winds as a possible origin of high-energy emission in LLAGNs, motivated by the LHAASO detection of NGC~4278. 
In our model, CRs accelerated at a wind-driven shock interact with the surrounding electromagnetic field and gas, and produce multimessenger signals including photons and neutrinos (Sections~\ref{sec:model dynamics} and \ref{sec:model emission}). 
Applying the model to NGC~4278, we found that our RIAF-wind model can account for the
observed TeV flux in the quiet period with comparable wind powers of
$\dot{K}_\mathrm{w}=1.5\times10^{43}~\mathrm{erg~s^{-1}}$ in a leptohadronic scenario
(Section~\ref{sec:leptohadronic}) and $\dot{K}_\mathrm{w}=1.7\times10^{43}~\mathrm{erg~s^{-1}}$
in a pure hadronic scenario (Section~\ref{sec:hadronic}).
We note that the leptohadronic scenario predicts a higher GeV flux than the inferred upper
limits from \textit{Fermi}-LAT observations \citep{VERITAS_2026} while the hadronic scenario
is consistent with the upper limits.
Based on this, we conclude that the current multiwavelength observations favor the hadronic
scenario and disfavor the leptohadronic one, although the \textit{Fermi}-LAT analyses of this
source have been mutually inconsistent (Section~\ref{sec:test}).
Using simple accretion arguments, we showed that such wind powers are achievable if the RIAF
is predominantly fed by ionized gas observed on $\sim 100$~pc scales through Bondi accretion
with a shallow accretion profile (Section~\ref{sec:gas accretion}).
Our scenarios will be tested from multiple perspectives in the future.
Radio imaging of the shocked region and TeV gamma-ray spectroscopy will provide direct tests
of the wind-emission scenario, while MeV observations could constrain competing jet models.
The accompanying neutrino flux, however, falls below the sensitivity of planned TeV--PeV detectors (Section~\ref{sec:test}).
We inferred that NGC~4278 is unlikely to be representative of LLAGNs in the neutrino sky, but an LLAGN population itself may provide a subdominant contribution to the extragalactic neutrino background if such RIAF-wind emission is common among this source class (Section~\ref{sec:DNB}).
Furthermore, our model also reproduced the TeV emission around Sgr~A* with a reasonable parameter set (Section~\ref{sec:Sgr A*}).

Investigating LLAGNs in the future is important for understanding the origin of the diffuse gamma-ray and neutrino backgrounds.
Further observational studies of several gas phases in LLAGNs will help to better constrain wind powers and gas densities, both of which are necessary for determining a \textit{pp}-induced gamma-ray luminosity.
In addition, coordinated gamma-ray, radio, and neutrino observations of NGC~4278 and similar LLAGNs will help to test the RIAF wind scenario more stringently.
A more detailed model that takes an AGN luminosity as an input and calculates gamma-ray and neutrino emission is required for a more precise estimation of the contribution to the extragalactic neutrino background, where a luminosity integration is conducted.
Such studies will quantify how much an LLAGN population, which accounts for $\sim$30\% of nearby galaxies, contributes to the diffuse gamma-ray and neutrino backgrounds.

\begin{acknowledgments}

The authors thank Yoshiyuki Inoue, Todd Thompson, and Tim Linden for useful comments and discussions. 
This research has made use of the NASA/IPAC Extragalactic Database (NED), which is funded by the National Aeronautics and Space Administration (NASA) and operated by the California Institute of Technology. 
N.S. was supported by JST SPRING, Grant Number JPMJSP2138, and is supported by JSPS Grant-in-Aid for JSPS Fellows Grant Number JP26KJ1621.
J.F.B. was supported by National Science Foundation Grant No.\ PHY-2310018.
The work of K.M. is supported by the NSF Grant No.~AST-2308021.
\end{acknowledgments}

\software{AMES \citep{2023MNRAS.524...76Z,Murase2024},
astropy \citep{2013A&A...558A..33A,2018AJ....156..123A,2022ApJ...935..167A},
matplotlib \citep{2007CSE.....9...90H},
numpy \citep{2020Natur.585..357H}
          }

\bibliography{TeV-NGC4278}{}
\bibliographystyle{aasjournalv7}

\end{document}